# Stability of clinical prediction models developed using statistical or machine learning methods

Richard D Riley[1*], Gary S Collins[2]

**Contact details:**

* corresponding author:

Professor of Biostatistics

e- mail: r.d.riley@bham.ac.uk

@Richard_D_Riley

[1] Institute of Applied Health Research, College of Medical and Dental Sciences, University of Birmingham, Birmingham, UK. B15 2TT.

[2] Centre for Statistics in Medicine, Nuffield Department of Orthopaedics, Rheumatology and Musculoskeletal Sciences, University of Oxford, Oxford, UK. OX3 7LD.

**Running title**: Stability of clinical prediction models

**Funding**: This paper presents independent research supported (for RDR and GSC) by the MRC-NIHR Better Methods Better Research panel (grant reference: MR/V038168/1), and (for RDR) the NIHR Birmingham Biomedical Research Centre at the University Hospitals Birmingham NHS Foundation Trust and the University of Birmingham. The views expressed are those of the author(s) and not necessarily those of the NHS, the NIHR or the Department of Health and Social Care. GSC was also supported by Cancer Research UK (programme grant: C49297/A27294).

**Word count**: 7100

# Abstract

Clinical prediction models estimate an individual's risk of a particular health outcome, conditional on their values of multiple predictors. A developed model is a consequence of the development dataset and the chosen model building strategy, including the sample size, number of predictors and analysis method (e.g., regression or machine learning; penalization and variable selection approach). Here, we raise the concern that many models are developed using small datasets that lead to *instability* in the model and its predictions (estimated risks). We define four levels of model stability in estimated risks moving from the overall mean to the individual level. Then, through simulation and case studies of statistical and machine learning approaches, we show instability in a model's estimated risks is often considerable, and ultimately manifests itself as miscalibration of predictions in new data. Therefore, we recommend researchers should always examine instability at the model development stage and propose instability plots and measures to do so. This entails repeating the model building steps (those used in the development of the original prediction model) in each of multiple (e.g., 1000) bootstrap samples, to produce multiple bootstrap models, and then deriving (i) a *prediction instability plot* of bootstrap model predictions (y-axis) versus original model predictions (x-axis), (ii) a *calibration instability plot* showing calibration curves for the bootstrap models in the original sample; and (iii) the *instability index*, which is the mean absolute difference between individuals' original and bootstrap model predictions. A case study is used to illustrate how these instability assessments help reassure (or not) whether model predictions are likely to be reliable (or not), whilst also informing a model's critical appraisal (risk of bias rating), fairness assessment (e.g. by comparing different ethnic groups) and further validation requirements. Stata and R code are also provided.

# 1   Introduction

Clinical prediction models are used to inform diagnosis and prognosis in healthcare,[1-3] and examples include EuroSCORE[4,5] and the Nottingham Prognostic Index.[6,7] They allow health professionals to predict an individual's outcome value, or an individual's risk of an outcome being present (diagnostic prediction model) or will occur in the future (prognostic prediction model). Prediction models may be developed using statistical or machine learning approaches, such as regression (e.g. logistic or Cox regression, and penalised adaptations such as the LASSO, elastic net or ridge regression) or machine learning methods (e.g. random forests, neural networks). These produce an equation or software object ('black box') to estimate an individual's outcome value or outcome risk conditional on their values of multiple predictors, which may include basic characteristics (such as age, weight, family history and comorbidities), biological measurements (such as blood pressure and biomarkers), and imaging or other test results.

A developed model is a consequence of the sample of data used to develop it, the predictors considered and the analysis approach, including (but not limited to) the model building framework (e.g. regression, random forests, neural networks), the use of any shrinkage or penalization methods, and the handling of missing values. The accuracy of predictions from the model depends on such facets. Assuming relevant predictors are available at the time of modelling, predictions are more likely to be accurate in new data when the development data is large, when the potential for overfitting is kept low, and when shrinkage or penalization methods are applied. Conversely, predictions are more likely to be unreliable when the development data is small (containing too few outcome events), when the model complexity (e.g., number of predictor parameters considered) is large relative to the number of outcome events, and when the modelling approach does not adjust overfitting. Unfortunately, in practice, many prediction models are developed using such approaches,[8-10] and so a concern is that model predictions may be unreliable and not fit for purpose. At the model development stage, this problem manifests itself as *model instability (*or *volatility) –* that is, the developed model (e.g. regression equation, random forest, neural network) may be very different if it were developed in a different sample of the same size from the same population. For example, depending on the modelling strategy and sample size, there may be volatility in the set of selected predictors, the weights assigned to predictors, the functional forms of predictors, the selected interactions, and so forth.

Such volatility leads to instability in model predictions, such that estimated risks depend heavily on the particular sample used and chosen modelling strategy. The larger the instability in model predictions, the greater the threat that the developed model has poor internal validity (in the development population) let alone external validity (in different populations). For this reason, stability checks should become a routine part of any research developing a new clinical prediction model. To promote this, here we use a case study and simulation to demonstrate the concept of instability, and then explain how to undertake stability checks at the model development stage using bootstrapping.

The article outline is as follows. In Section 2, we provide an overview of previous papers examining instability, and then define four levels of stability in model predictions, with illustration using a simulation study. Section 3 proposes how instability can be examined at the model development stage itself, using bootstrapping and the presentation of instability plots and measures. Section 4 illustrates the methods for various case studies and modelling approaches, and shows the extent of instability in an individual's estimated risk. Section 5 extends to the use of instability to examine fairness, clinical utility and classification. Section 6 concludes with discussion. The work is based on lectures held in the series "Education for Statistics in Practice" by Richard Riley and Gary Collins at the March 2022 conference of the "Deutsche Arbeitsgemeinschaft Statistik" in Hamburg, Germany, where instability was discussed alongside related issues of sample size, reporting and critical appraisal. Slides are available at http://www.biometrische-gesellschaft.de/arbeitsgruppen/weiterbildung/education-for-statistics-in-practice.html.

## 2 Defining stability of clinical prediction models

To set the scene, we briefly discuss previous work and then define four levels of stability in risk estimates from a clinical prediction model.

### 2.1 Previous research

An early investigation of stability in prediction models is the work of Altman and Andersen,[11] who use bootstrapping to investigate the instability of a stepwise Cox proportional hazards regression model, in terms of the variables selected and, more importantly, the model's predictive ability. They construct bootstrap confidence intervals for each individual's estimated risk and show that such intervals are markedly wider than when derived solely on the original model. Other studies have also used bootstrapping to highlight the instability of variable selection methods and non-linear relationships,[3,12-15] and stress how in small samples the selection of predictors and their functional forms are highly unstable, and so should be considered with caution. Harrell stresses the importance of pre-specifying modelling decisions (e.g. placement of knot positions in a spline function), as otherwise there is greater potential for instability in the final model produced.[16]

Moreover, instability also depends on the number of candidate predictor parameters considered, and so Riley et al. and van Smeden et al. suggest sample size calculations for model development (of regression based prediction models) to limit the number of candidate predictor parameters relative to the total sample size and number of events,[17-21] though stability might be improved by using prior information[22] and use of penalisation methods. However, studies have shown that even penalisation methods such as LASSO and elastic net are unstable in their selection of predictors (due to uncertainty in estimation of the tuning parameters), especially in small sample sizes where they are arguably most needed, leading to miscalibration of predictions in new data.[23-25] For this reason, Martin et al. recommend model developers should use bootstrapping to investigate the uncertainty in their model's penalty terms (shrinkage factors) and predictive performance[26]. Others studies have emphasised using methods to improve stability in the penalisation approach,[27] such as repeat *k*-fold cross-validation to estimate penalty factors,[24] or ensemble methods that incorporate boosting or bagging (e.g. XGBoost, random forests).[28] Yet, even with a reasonable sample size and appropriate modelling methods, Pate et al. use a resampling study to demonstrate the (often huge) instability of individualised risk estimates from prediction models of cardiovascular disease,[29] and we build from this in Section 3.

## 2.2 Levels of stability in model predictions

In the remainder of this paper, we focus on instability of model predictions (rather than instability in the model specification or selection of predictors and their functional form), akin to Altman and Andersen,[11] and Pate et al.[29] We focus on quantifying and measuring (in)stability of predictions produced by one particular model building strategy, in terms of the extent to which predictions for an individual may differ depending on the development sample used. We are not interested in instability of differences in predictions between two competing model strategies. No single model is ever 'correct', but our premise is that researchers should at least aim to produce a model that is internally valid (for the particular model development approach taken) and stability checks help examine this.

We begin by framing prediction stability in a hierarchy of four levels of stability of estimated risks defined by: (1) the mean, (2) the distribution, (3) subgroups, and (4) individuals. The four levels move from stability at the population level (1 and 2) to then groups (e.g. defined by ethnicity) and individuals (e.g. defined by values of multiple predictors). As clinical prediction models typically aim to inform individualised decision making, level (4) will typically be the most important.

To illustrate these four levels, we will use a simulated example where the set-up is as follows:

- A population has a true overall risk of 0.5 of an outcome event
- An individual's true logit event risk can be expressed by a single predictor, $X$, distributed normally with mean 0 and standard deviation of 2.
- Choose the sample size ($n$) for evaluation (in our results below, we focus mainly on $n$ = 100, but also consider 50, 385, 500, 1000 and 5000)
- For a random sample of $n$ individuals, values of $X$ are drawn for each individual from $N(0,4)$ and values of binary outcome $Y$ are drawn from $\text{Bernoulli}(p_i)$ where $\text{logit}(p_i) = \text{LP} = X$ (i.e. a logistic regression model with intercept zero and a regression coefficient for $X$ of 1). Values of 10 noise variables ($Z_1$ to $Z_{10}$) are also each drawn from $N(0,1)$.
- Using the datasets of $n$ individuals, a prediction model is developed using logistic regression with a LASSO penalty (implemented using 10-fold cross-validation and the tuning parameter chosen as lambda.min) considering the 11 candidate predictors ($X$ and $Z_1$ to $Z_{10}$).
- The fitted model is then used to make predictions in 100000 other individuals from the same population (with their $X$ and $Z_1$ to $Z_{10}$ values generated using exactly the same process as used for the development data);
- The previous steps are repeated 1000 times: that is, on each occasion generate a random sample of $n$ individuals for model development, develop the model using logistic regression with a LASSO penalty considering the 11 candidate predictors, and then apply the fitted model to calculate estimated risks in the same 100000 individuals.

Hence, this process generates 1000 different models (each distinct, with their own set of included predictors and magnitude of predictor effects), and each of the 100000 individuals has 1000 estimates of risk (one for each model). Of course, in practice, a model developer only produces one model. However, we repeat the process 1000 times to stress that any one model is just an 'example model' (a phrase coined by Prof Frank Harrell Jr, e.g. https://www.fharrell.com/post/split-val/) and different models would have been developed had a different dataset (of the same sample size) been obtained from the development population. The concept of 'example models' should motivate prediction model developers to always ask: is there instability in predictions from the prediction model they have just developed relative to predictions from all the other potential example models?

### 2.2.1 Level 1: Stability in a model's mean estimated risk

The first stability level, and the bare minimum requirement, relates to the developed model's mean estimated risk. To demonstrate this, consider the variability in the mean estimated risk from 1000 models for various model development sample sizes (**Figure 1**(a)), when the model is applied to a sample of 10000 individuals. With a model development sample size of 100 participants, 95% of the models' mean estimated risks are between about 0.42 and 0.58. With a sample size of 5000 participants, the 95% range is much narrower (0.49 to 0.51), whilst with just 50 participants it is much wider (0.36 and 0.64). Hence, the smaller the development dataset, the greater the instability in a model's mean estimated risk, and the likely downstream consequence is miscalibration between the mean estimated and mean observed risk in the population (also known as miscalibration-in-the-large). For this reason, we have previously suggested it is essential to ensure the minimum sample size for model development will estimate the overall risk precisely (e.g., within 0.05 of the true value),[18][19] which is 385 participants for this particular example.

### 2.2.2 Level 2: Stability in a model's distribution of estimated risks

Moving beyond the mean estimated risk, stability of the entire distribution of estimated risks is the next level to be evaluated. The smaller the model development sample size, the more variable the shape of the distribution can be. For example, **Figure 1**(b) shows the distribution of estimated risks (in the population of 100000 individuals) is noticeably different for two example models developed using a sample size of 100 participants.

The downstream consequence of such large level 2 instability is that an example model's estimated risks will likely be miscalibrated with observed risks in the population, as evident from examining calibration across the whole spectrum of estimated risks from 0 to 1. This can be shown using calibration plots of estimated versus observed risks, and by examining the potential deviation of flexible calibration curves from the 45-degree line of perfect agreement. For example, Figure 2 shows calibration curves in the population (100000 individuals) for 1000 example models developed using sample sizes from 100 to 5000. Variability in the curves increases as the sample size decreases, and in the small sample sizes there is quite substantial spread, indicating the potential for any one example model to be miscalibrated in new data. However, the curves become more stable when at least the minimum sample size of 385 is achieved. A further example of instability of calibration curves is shown in Supplementary Figure S1, for a situation where a larger minimum sample size is required, and so the instability of curves at sample sizes of 100 and 200 is even more pronounced.

**Figure 1** Instability of estimated risks from applying example prediction models (developed using a particular sample size) to the same population of 100000 individuals: each example model was produced from a logistic regression with a LASSO penalty fitted to a different random sample of individuals from a population with a true overall risk of 0.5, considering 1 genuine predictor ($X \sim N(0,4)$) and 10 noise variables ($Z_1, \ldots, Z_{10} \sim N(0,1)$).

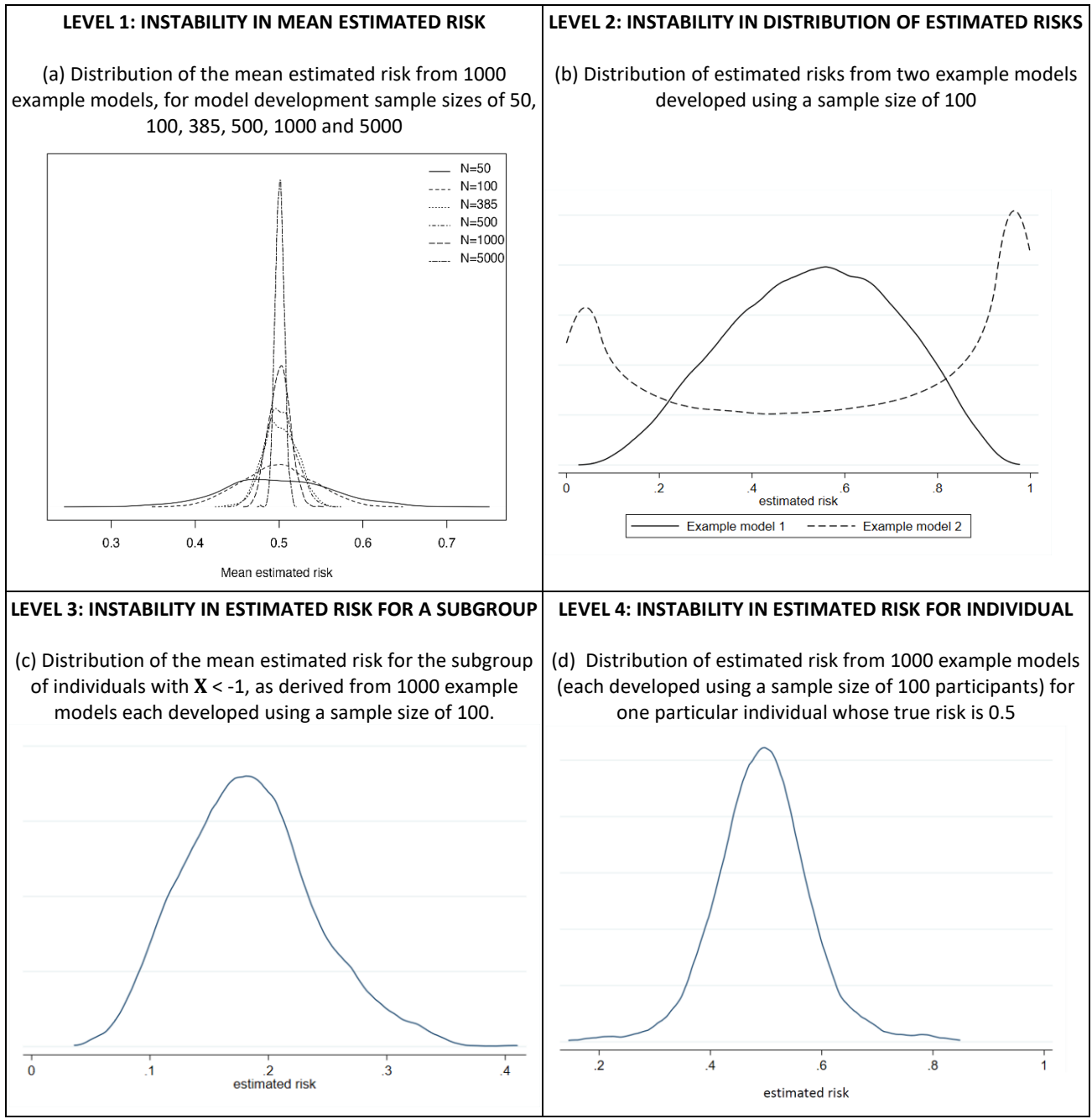

**Figure 2** Each plot shows instability of calibration curves for 1000 example models developed using a particular sample size (100, 200, 385, 500, 1000 or 5000) when each model is applied to the same population of 100000 individuals: each example model was produced from a logistic regression (LR) with a LASSO penalty fitted to a different random sample of individuals from a population with a true overall risk of 0.5, considering 1 genuine predictor (X~N(0,4)) and 10 noise variables (Z1, …, Z10 ~ N(0,1))

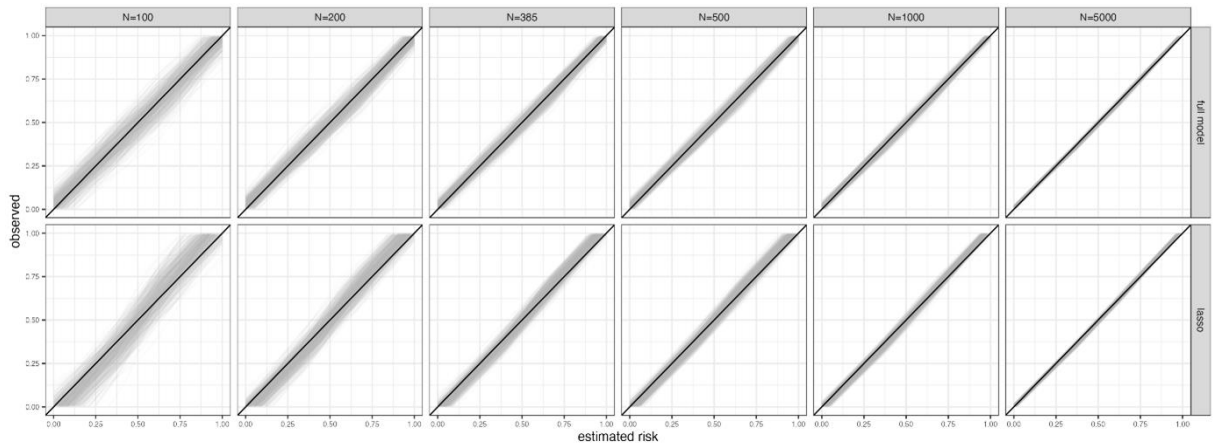

Another way to summarise the overall agreement between observed and predicted values is using the mean absolute prediction error (MAPE), which is the mean (over all individuals) of the absolute difference between estimated risks (from the developed model) and the true risks (as defined by the data generating model in the simulation study). For example, returning to our simulated data example, Figure 3 shows MAPE in the population for 1000 example models developed using sample sizes from 50 to 5000. Variability in MAPE increases as the sample size decreases, and is considerable in the very small samples, but starts to stabilise from the minimum sample size of 385 participants. Figure 3 also shows that the average MAPE is higher for an unpenalised logistic regression model (a 'full model' forcing all predictors in regardless and without shrinkage of predictor effects) compared to using a logistic regression with a LASSO penalty, emphasising why LASSO is often preferred. However, the instability in MAPE is generally larger for the LASSO. Similarly, the instability in calibration curves for the LASSO is slightly larger than the full model (Figure 2), especially at the lower sample sizes.

Level 2 instability in a model's distribution of risks also leads to instability in the model's discrimination performance, for example as measured by the C-statistic (also known as the area under the curve for binary outcomes). We return to this in Section 5.2.

**Figure 3 Mean absolute predictor error (MAPE) in a large population of 100000 individuals for 1000 example models developed using a logistic regression with all predictors forced in ('full model') or with a LASSO penalty, for development sample sizes of 50, 100, 385, 500, 1000 and 5000**

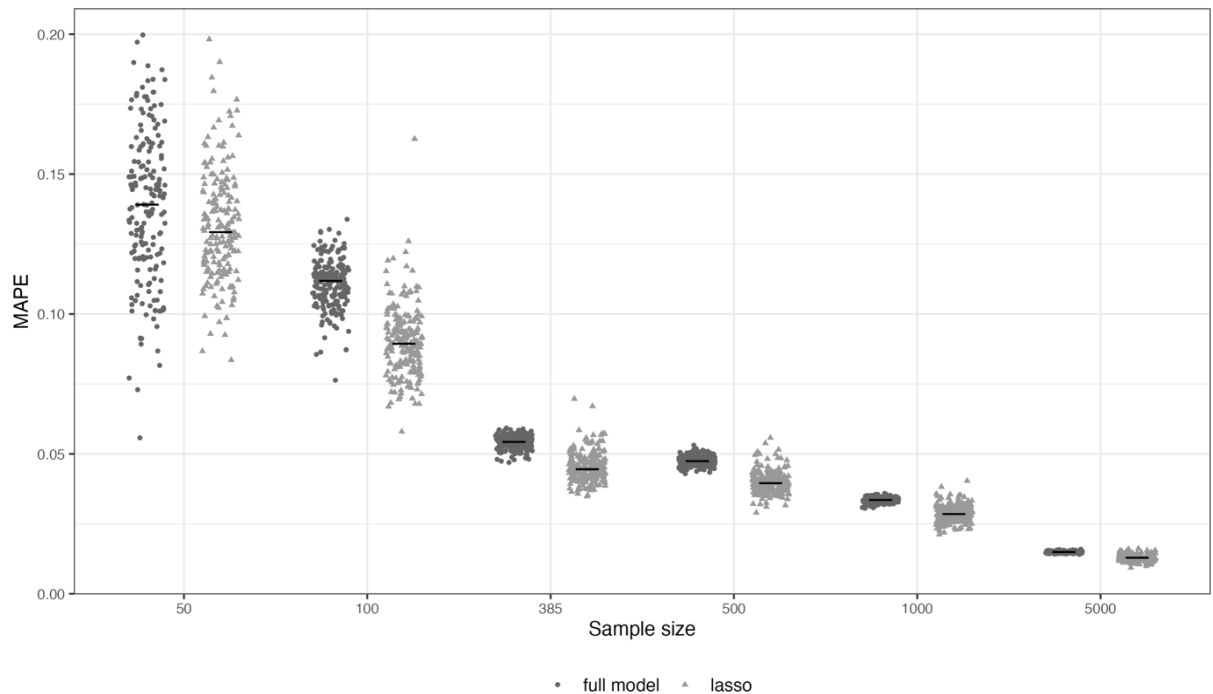

### 2.2.3 Level 3: Stability in a model's predictions for subgroups

Even if a model's predictions appear stable at levels 1 and 2, there may still be instability in the predictions for subgroups defined by a particular covariate (which may or may not be a predictor in the model). For example, returning to our simulated example with a model development sample size of 100, let us consider the subgroup of individuals defined by an X value < -1. Figure 1(c) plots the distribution of the mean estimated risk for this subgroup from the 1000 example models and instability is reflected by large variability in predictions (mainly between 0.1 and 0.3). Thus, any two example models may differ considerably in their estimated risk for this subgroup. Stability of predictions for subgroups is also relevant when considering algorithmic fairness (see Section 5.1).

### 2.2.4 Level 4: Stability in a model's predictions for individuals

Lastly, stability of predictions should be evaluated at the individual level, as defined by values of multiple covariates (patterns of covariate values). Instability at the individual level is often severe, which is a huge concern for models aiming to guide clinical practice for individuals, as users (e.g., clinicians, patients) need some degree of confidence that an individual's estimated risk is reliable enough to have a role in their clinical decision making. Here we define 'reliable' as different example models producing similar estimated risk risks for an individual; the more dissimilar the estimated risks, the more unreliable are the estimated risks from the model.

For our simulated example with a development sample size of 100, **Figure 1**(d) shows the distribution of estimated risks from 1000 example models applied to one particular individual whose true risk is about 0.5. Even though the individual's predictor values do not change (as it is the same individual), the predictions vary hugely across the example models, with minimum and maximum estimated risks of 0.14 and 0.85, respectively. Hence, this individual's estimated risk from one example model is hugely unstable and so unreliable. Subsequently there may be large instability in clinical decisions (e.g. based on clinical decision thresholds, see Section 5.2) and risk communication (e.g. for shared decision making) for that individual, thus having the potential for harm.

More broadly, Figure 4 shows variability of estimated risks for nine individuals with true risks (defined by the data generating model) between 0.1 and 0.9, across 1000 example models developed using sample sizes from 50 to 5000 participants. With 5000 participants, there is only small variability in the estimated risks across the example models for each individual, though the individual with a true risk of 0.5 still varies anywhere between about 0.1 of the true value. As the sample size for model development decreases, the variability increases. In particular, at the smallest sample sizes of 50 and 100 participants, the volatility is enormous, with estimated risks spanning the entire probability interval from 0 to 1 for most individuals highlighting large instability concerns and that the model is unreliable (different example models would lead to different clinical decisions or discussions about patient reassurance or likely outcomes). Even at the minimum sample size of 385 participants (which recall showed low instability in MAPE and calibration curves; see Figure 2 and Figure 3) the instability is quite remarkable, emphasising how the minimum sample size calculation targets level 1 and 2 stability, but not necessarily levels 3 and 4.

**Figure 4** Instability of estimated risks across 1000 example prediction models for each of nine individuals with true risks between 0.1 and 0.9, for model development sample sizes ($n_D$) of 50, 100, 385, 500, 1000 and 5000 participants: each example model was produced from a logistic regression (LR) with a LASSO penalty fitted to a different random sample of individuals from a population with a true overall risk of 0.5, considering 1 genuine predictor ($X \sim N(0,4)$) and 10 noise variables ($Z_1, …, Z_{10} \sim N(0,1)$)

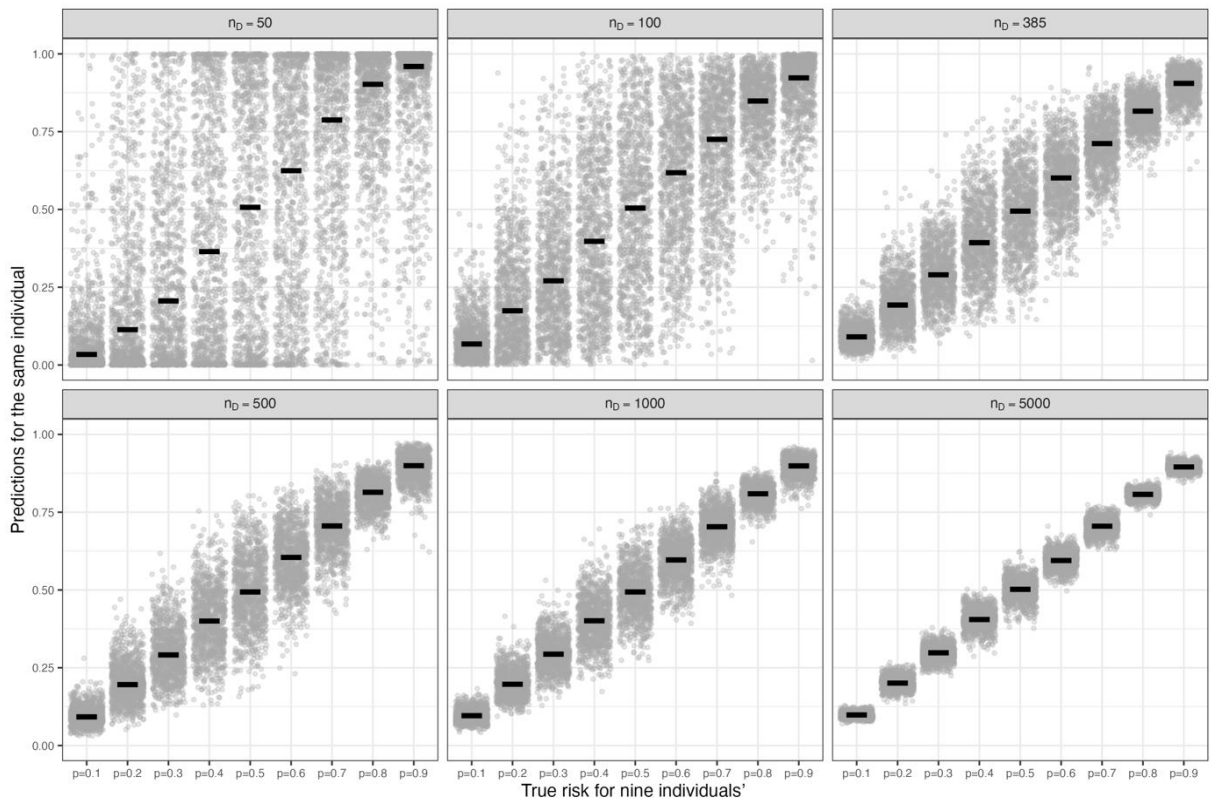

# 3 Quantifying instability in model development studies

In Section 2 our simulated example examined instability by comparing estimated risks to true risks for each individual. However, in practice when developing a model researchers will not know the 'true' risk of each individual, and so need to examine instability in a different way, using the development data itself. Building on earlier work,[11][29] we now outline a bootstrap process to do this and show how to derive various instability plots and measures.

## 3.1 Bootstrap assessment of instability

Bootstrapping is a technique to examine sampling variability and so provides a natural way to examine the instability of predictions from a developed model using a particular dataset. Using the model development dataset of $N$ participants, the required bootstrap process is explained in Box 1. Instability is a reflection of the sample size, so it is essential that each bootstrap sample is the same sample size as the original model development dataset. Similarly, instability is a consequence of the model development approach, and so the prediction models generated using the bootstrap samples in step 4 must use the same model development approach as taken originally. So, for example, the same model specification (e.g. logistic regression), any tuning parameter estimation (e.g. cross-validation), and variable selection method (e.g. backwards elimination, penalisation, etc) should be used. Some parts of the model building process (e.g. selection of non-linear functions for continuous predictors) may not be easily implemented (automatically) in each bootstrap sample (e.g. particularly if multiple imputation is being used to handle missing values) and in these instances, some compromise may be required to make the approach practical.

## 3.2 Numerical summaries and graphical presentations of instability

The predictions from step 6 can be used to derive instability plots and measures, as follows.

### 3.2.1 Prediction instability plot

The instability of a prediction model is reflected by the variability and range of individual-level predictions ($\hat{p}_{bi}$) from the $B$ bootstrap models. This can be shown graphically in a *prediction instability plot*, which is a scatter of the $B$ predicted values (y-axis) for each individual against their original predicted value (x-axis). This plot will typically include extreme values, and so alternatively (or additionally) a 95% range could be presented for each individual, defined by the 2.5<sup>th</sup> and 97.5<sup>th</sup> percentile of their $\hat{p}_{bi}$ values. The lower and upper values can be smoothed across individuals using a loess curve,[30] to essentially form a 95% confidence interval band for the estimated risks from the bootstrap models.[11] The bandwidth for the smoothing process is subjective and example specific, but generally we found values between 0.2 and 0.8 to be sensible. Examples are given in Section 4.

**Box 1 The bootstrap process to examine instability of predictions from a clinical prediction model after model development**

> Context: a prediction model has just been developed using a particular model building strategy, and the model developers want to examine the potential instability of predictions from this model. To do this using the model development dataset of $n$ participants, we recommend the following bootstrap process is applied:
>
> - Step 1: Use the developed model to make predictions ($\hat{p}_i$) for each individual participant ($i = 1$ to $n$) in the development dataset.
> - Step 2: Generate a bootstrap sample with replacement, ensuring the same size ($n$) as the model development dataset.
> - Step 3: Develop a bootstrap prediction model in the bootstrap sample, replicating exactly (or as far as practically possible) the same model building strategy as used originally.
> - Step 4: Use the bootstrap model developed in step 3 to make predictions for each individual ($i$) in the original dataset. We refer to these predictions as $\hat{p}_{bi}$, where $b$ indicates which bootstrap sample the model was generated in ($b$ = 1 to $B$).
> - Step 5: Repeat steps 2 to 4 a total of ($B - 1$) times, and we suggest $B$ is at least 200.
> - Step 6: Store all the predictions from the $B$ iterations of steps 2 to 5 in a single dataset, containing for each individual a prediction ($\hat{p}_i$) from the original model and $B$ predictions ($\hat{p}_{1i}, \hat{p}_{2i}, \ldots, \hat{p}_{Bi}$) from the bootstrap models.
> - Step 7: Summarise the instability in model predictions, using prediction instability plots, calibration instability plots and the mean absolute predictor error (MAPE), see main text.

### 3.2.2   Calibration instability plot

When a calibration curve is estimated in the model development dataset, the model's predictions may look well calibrated (as the model was fitted in that dataset) when actually they are poorly calibrated in the population. This concern may be exposed by examining instability in the calibration curves for the $B$ bootstrap models when assessed in the original dataset. The $B$ curves are overlayed on the same plot, together with the original calibration curve of the original model applied in the original data. We refer to this as a *calibration instability plot*. The wider the spread of the $B$ calibration curves, the greater the instability concern (and thus the threat of model miscalibration in the actual population). With many curves the plot may be dense and unclear, and so displaying a random sample of 200 will often suffice, or opacity of the lines could be changed (e.g., changing the *alpha* aesthetics value in ggplot2 in R). Examples are given in Section 4.

### 3.2.3   Mean absolute predictor error (MAPE)

For each individual, the *mean absolute predictor error* (MAPE) can be calculated as the mean absolute difference between the bootstrap model predictions and the original model prediction; that is,

$$\text{MAPE for individual } i = \frac{\sum_{b=1}^{B}|\hat{p}_{bi} - \hat{p}_i|}{B}$$

Though we do not know an individual's true risk, we still refer to 'error' as the differences in the original and bootstrap predictions are essentially calculating what the error would be if the bootstrap models were the truth. The *average MAPE* across all individuals can be summarized using:

$$\text{average MAPE} = \frac{\sum_{b=1}^{B}\sum_{i=1}^{N}|\hat{p}_{bi} - \hat{p}_i|}{BN}$$

Whilst this single measure is a useful overall summary, by aggregating across individuals it masks potentially larger (or smaller) differences in instability at the individual level or within particular regions of risk. Hence, the average MAPE should always be accompanied by an instability plot of individual-level MAPE values, as described next.

### 3.2.4   MAPE Instability plot

MAPE can be shown graphically in a *MAPE instability plot*, which is a scatter of the MAPE value (y-axis) for each individual against their estimated risk from the original prediction model (x-axis). This plot reveals the range of MAPE values and helps identify if and where instability is of most concern for the original predictions. Examples are given in Section 4.

# 4 Investigation of stability in various case studies

We now use case studies to illustrate our proposed bootstrap approach and the instability plots and measures. We consider various model development methods and different sample sizes. Our aim is to illustrate how researchers can check stability after developing a clinical prediction model. We do not aim to identify the 'best' modelling technique per se.

In all our case studies, models are developed using the GUSTO-I dataset that contains individual participant level information on 30-day mortality following an acute myocardial infarction. The aim is to develop a prediction model for risk of death by 30 days. The dataset is freely available, for which we kindly acknowledge Duke Clinical Research Institute,[31] and can be installed in R by typing: load(url('https://hbiostat.org/data/gusto.rda')). In the full dataset, there are 40830 participants with 2851 deaths by 30 days, and thus the overall risk is about 0.07. In some case studies we reduce the sample size by taking a random subset of participants; this allows us to examine how instability changes according to the sample size for model development. Seven predictors are considered for inclusion to predict 30-day mortality: Sex (0=male, 1=female), Age (years), Hypertension (0=no, 1=yes), Hypotension (0=no, 1=yes), Tachycardia (0=no, 1=yes), Previous Myocardial Infarction (0=no, 1=yes), and ST Elevation on ECG (number of leads). For reference, fitting a logistic regression with no penalisation or predictor selection using the full dataset and these seven predictors leads to a $C$ statistic of 0.8, and Nagelkerke $R^2$ of 0.2 (20% explained variation).

## 4.1 Unpenalised logistic regression forcing in 7 predictors

Our first case study involves fitting a standard (unpenalised) logistic regression model forcing in the aforementioned 7 predictors, for each of 'large' and 'small' sample size scenarios. The large sample size is the full dataset of 40830 participants (2851 deaths), which corresponds to an Events per Predictor Parameter (EPP) of 407. We defined the small sample size scenario as 300 participants (21 deaths), corresponding to an events per predictor parameter (EPP) of 3. For the small sample size scenarios, we sampled randomly a subset from the entire GUSTO I data. Figure 5 shows the instability plots and measures for both scenarios. It is clear that instability in individual predictions is very low in the large sample size scenario, with an average MAPE of 0.0027, indicating that on average across individuals the absolute difference in the developed model's predictions and the bootstrap model's predictions is just 0.0027. There is low variability in calibration curves and individual risk predictions across bootstrap samples. Given this, the model developed in the full dataset has strong potential to perform well in the target population (assuming the data sampled reflects the target population).

**Figure 5 Instability plots and measures for a prediction model developed using unpenalised logistic regression forcing in seven predictors, in each of three different sample size scenarios**

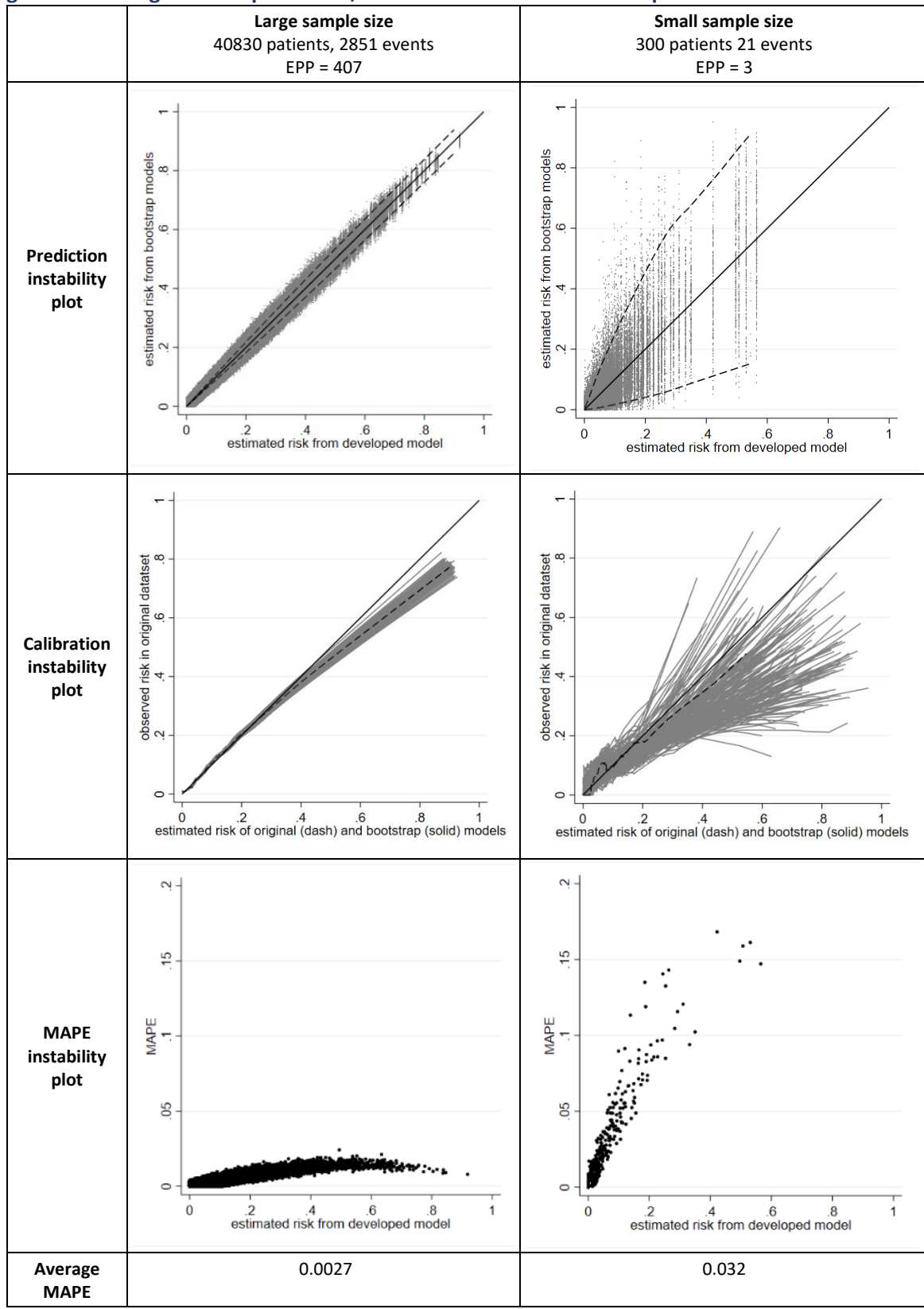

However, instability concerns are more substantial in the small sample size scenario, with an average MAPE of 0.03. Whilst the MAPE instability plot shows a maximum of only 0.02 in the large sample size setting, it is 0.17 in the small scenario, and MAPE values exceed 0.05 for original estimated risks of 0.05 or higher. The prediction and calibration instability plots also reveal instability in the small scenario. For example, for an individual with an original estimated risk of 0.5, the uncertainty in estimated risks from the bootstrap models ranges from 0.1 to 0.8, and thus spans low risk to high risk; this is clearly unhelpful, and casts doubt on the reliability of the individual risk estimates. Calibration curves also deviate substantially from the 45-degree line. Thus, in the small sample size setting, it is quite unlikely that the model will be reliable in the target population.

## 4.2 Situations with many noise variables and the LASSO

There are many examples in the literature where the number of candidate predictors (and predictor parameters) considered for inclusion in a prediction model is large relative to the sample size (and in extreme cases such as EPP << 1, this is often referred to as high dimensional). Often penalisation methods, such as the LASSO, are heralded in this situation to resolve the problem of small sample sizes and low EPP. However, penalisation methods are not carte blanche in this situation.[23][24] To illustrate this, we now consider developing a model with 27 potential predictor parameters, involving the 7 previous predictors and an additional 20 noise variables (randomly generated from a N(0,1) distribution with no association with the outcome). We consider a small sample size of 752 participants (53 deaths), and thus an EPP of about 2, and develop two different models using,

(a) logistic regression forcing all 27 predictors to be included, and
(b) logistic regression with a LASSO penalty, using 10-fold cross-validation to estimate the tuning parameter.

The corresponding instability plots and measures are shown in Figure 6. Approach (a) has the most instability, with many individuals having a wide span of predictions from the bootstrap models. For example, when forcing all 27 predictors into the model, those with an estimated risk of 0.3 from the original model have estimated risks ranging from about 0 to 0.9 from the bootstrap models. The LASSO (approach (b)) reduces the instability (average MAPE reduced from 0.038 to 0.029), due to the shrinkage of predictor effects. Nevertheless, the instability is still considerable; for example, in individuals with an estimated risk of 0.2 from the original model, their estimated risks range from about 0.05 to 0.5 in the bootstrap models. There considerable spread in the calibration curves. Thus, though LASSO shows some improvement, both developed models are unstable and require further validation; indeed, the instability assessments suggest neither model is likely to perform well in new data, and illustrates again the concern of developing a clinical prediction model in small data.

**Figure 6 Instability plots and measures for a prediction model developed in a small sample size scenario (752 participants, 53 events, 27 candidate predictors, EPP of 2), for each of three different modelling approaches**

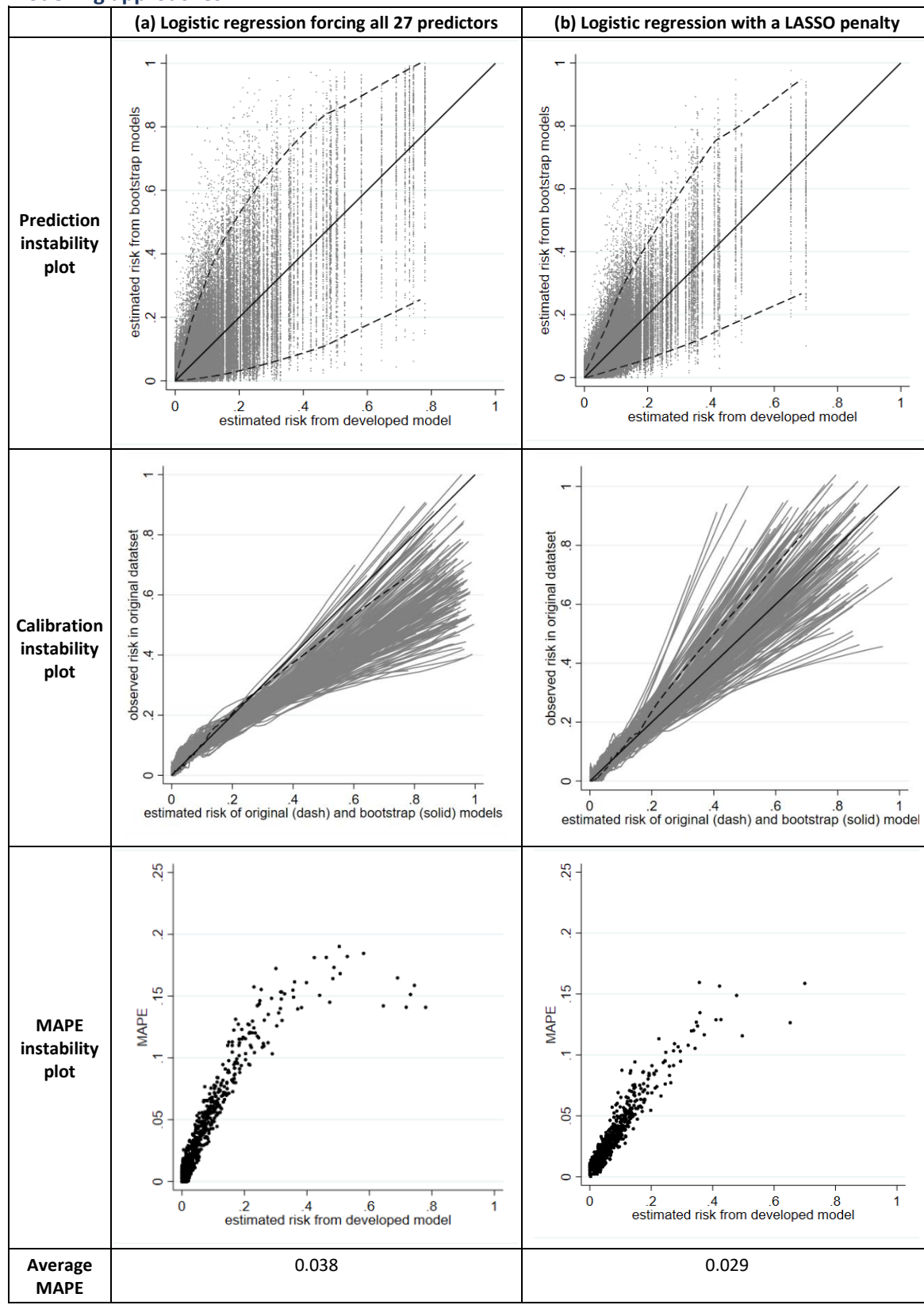

## 4.3 LASSO and uniform shrinkage with a minimum sample size for model development

We now examine instability in a dataset that meets our minimum sample size criteria,[18] which targets low overfitting and precise estimation of the overall risk. With 7 predictors, the minimum sample size scenario was defined by applying our sample size formulae assuming a Cox-Snell $R^2$ of 0.08 and targeting a uniform shrinkage factor of 0.9, leading to 752 participants (53 deaths) corresponding to an EPP of 7.5.[18,19] We develop two models in a random sample size of 752 participants, using:

(a) Logistic regression with a LASSO penalty, using 10-fold cross-validation to estimate the tuning parameter

(b) Logistic regression followed by a uniform shrinkage of predictor effects (estimated using the heuristic shrinkage of Van Houwelingen and Le Cessie[32]) and re-estimation of the model intercept to ensure calibration-in-the-large.

The instability results are very similar for both approaches (see Supplementary Figure S2), echoing previous work showing the choice of penalisation approach is less important as the sample size increases.[33] The average MAPE is 0.019 for LASSO and 0.018 for uniform shrinkage, with non-negligible variability in calibration curves and individual predictions across bootstrap samples (see Figure 7(a) for LASSO results). Most individuals have a MAPE < 0.1 but MAPE increases as the model's estimated risk increases, and the variability in bootstrap estimated risks is quite pronounced for some individuals. For example, individuals with an original estimated risk of 0.4 have a 95% range of about 0.2 to 0.6 in their estimated risks from bootstrapping. Thus, it is clear that further validation of these models (e.g., in an external validation study) is still required to be reassured that they are reliable.

**Figure 7 Instability plots and measures for LASSO and random forest models developed in a dataset of 752 participants (53 events) with 7 candidate predictors**

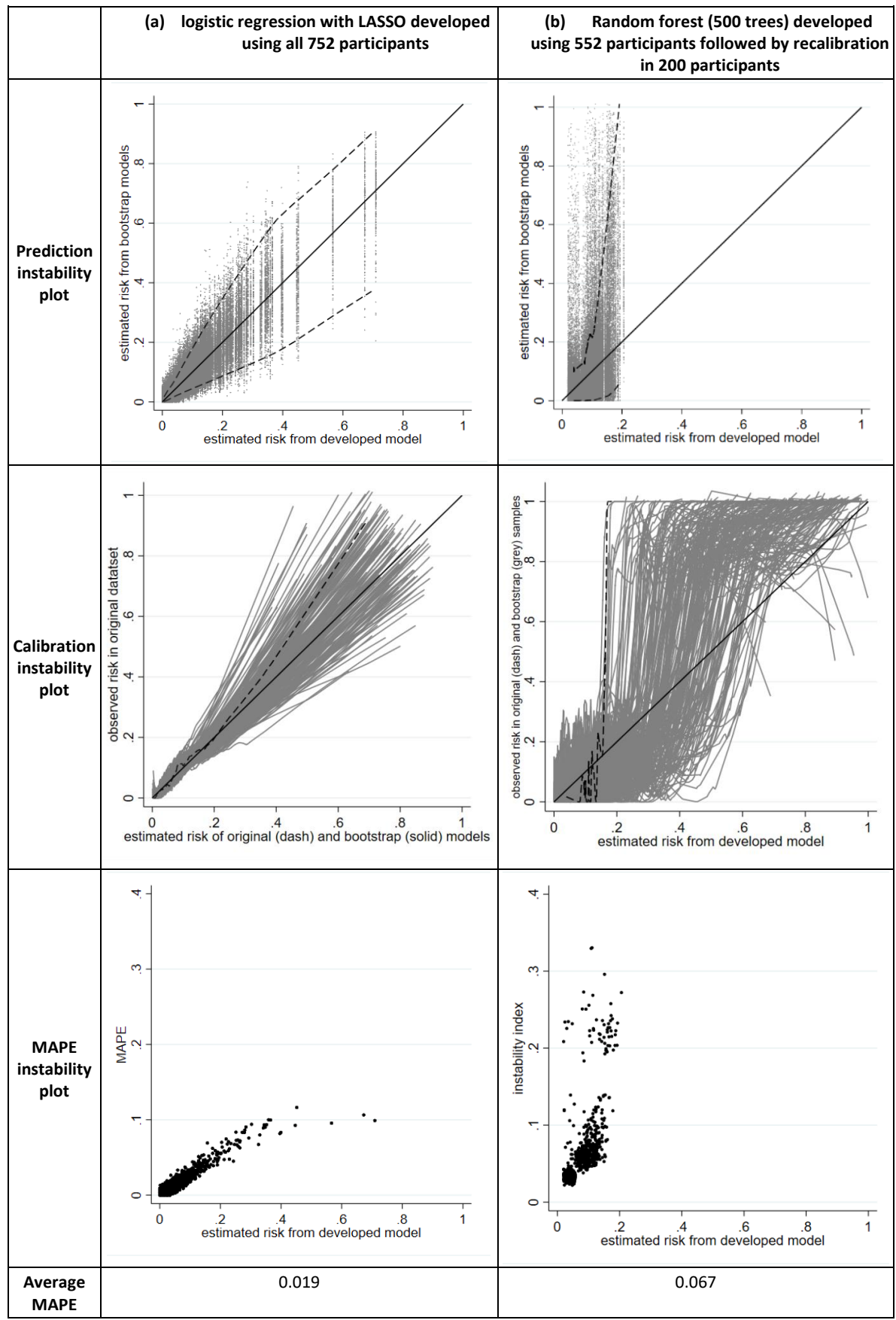

## 4.4 Random forest examples

The issue of instability also applies to modelling approaches other than (penalised) regression, such as random forests. To illustrate this, now we examine instability of a random forest developed using the same 752 participants (53 deaths) as the previous section, using 500 trees. Supplementary Figure S2(c) shows the instability is greater than when using LASSO and uniform shrinkage approaches, with an average MAPE of 0.044, and individual MAPE values often above 0.1 and even reaching 0.3. The instability in individual risk predictions is considerable, with an individual with original risk of 0.1 having bootstrap risks anywhere between 0 and 0.5, and those with original risk of 0.6 or above having bootstrap risks ranging from 0 to 1. Thus, it is clear that predictions based on this random forest are unstable and the model is unlikely to have reliable predictions in new data.

One aspect of random forests is that they often exhibit miscalibration.[34][35] For this reason, a hold-out dataset is often used to recalibrate by fitting a logistic regression model with the only covariate being the logit-risk estimates from the original random forest (this recalibration is sometimes referred to as Platt scaling[35]). To examine stability of this approach, let us consider randomly splitting the 752 participants into 452 for developing the random forest and 300 participants for the recalibration exercise. To properly assess instability in this situation, the bootstrap process must also include the random split element when producing bootstrap samples for the random forest and recalibration parts, in order to properly reflect the variability in the entire model building process. The instability results are shown in Figure 7(b), and there is still huge concern. The distribution of estimated risks is much narrower than originally, due to the recalibration, but the prediction instability index still shows a wide 95% range (e.g. between 0 and 1 at estimated risks of 0.2), and calibration curves show large variability (and unusual shapes) across bootstrap samples. Interestingly, the average MAPE is 0.067, which is larger than before recalibration (0.044). A key reason for this is that instability is increased by splitting the dataset, as this reduces the sample size (by 300 participants) to develop the random forest. Further, the recalibration sample size is also small, making it also hard to estimate the recalibration model precisely, and thus created further volatility in the model development process.

This whole exercise demonstrates the problem with using random forests in small sample sizes, which echoes previous research highlight the need for substantially larger sample sizes to reduce instability in random forest models.[36]

# 5 Further role of stability assessments

## 5.1 Informing fairness by examining stability in subgroups

There is increasingly recognition that prediction models need to be evaluated for important subgroups, especially to help ensure fairness and accuracy in under-represented or marginal groups such as defined by sex, ethnicity or deprivation.[37] Instability evaluations can play an important role in this context. For example, consider the scenario of Section 4.3, and whether the LASSO model would perform equally well in both males and females. Supplementary Figure S3 shows the instability plots are quite similar for males and females, with similar variability in calibration curves and individual risk predictions. The average MAPE is slightly higher for females than males (0.027 and 0.017), suggesting the model predictions may be slightly less reliable for females in new data. This may be due to predictions being slightly higher for females compared to males, as the predictor effect for sex corresponds to an odds ratio of 1.15 (females versus males), and higher estimated risks are likely to be more unstable than those closer to zero. Nevertheless, the stability differences between males and females are quite small, so there is no immediate concern that the model is unfair in regard to stability of predictions for males and females. In other situations, especially when some subgroups have substantially smaller sample sizes than others, stability differences may be apparent.

## 5.2 C-statistic

Instability in the distribution of a model's estimated risk will naturally lead to instability in the model's discrimination performance, and this can also be examined using the bootstrap process. The C-statistic should be calculated for each bootstrap model applied in the original dataset, and the greater the variability in C-statistic estimates, the greater the instability concern. For example, Supplementary Figure S4 shows the histogram of the C-statistic values obtained for the LASSO model from Section 4.3 applied in 1000 bootstrap samples, with values ranging from about 0.74 to 0.79.

## 5.3 Clinical utility

Where the goal is for predictions to direct decision making, a model should also be evaluated for its overall benefit on participant and healthcare outcomes; also known as its *clinical utility*.[38][39][40] For example, if a model estimates a patient's outcome risk is above a certain threshold value, then the patient and their healthcare professionals may decide on some clinical action (e.g. above current clinical care), such as administering a particular treatment, monitoring strategy, or life-style change. The clinical utility of this approach can be quantified by the *net benefit*, a measure that weighs the benefits (e.g., improved patient outcomes) against the harms (e.g., worse patient outcomes, additional costs).[41 42] It requires the researchers to choose a (range of) threshold(s), at or above which there will be a clinical action. A decision curve can be used to display a model's net benefit across the range of chosen threshold values.[41-43]

Therefore, it is important to also examine stability of decision curves, on a *decision curve instability plot*, which overlays the decision curves for the original model and the $B$ bootstrap models applied to the original dataset. For example, returning to the first case study (Section 4.1), the variability (instability) in decision curves from bootstrap models is tiny when the sample size is large (Figure 8). In the small sample size situation, variability is much more pronounced though the impact of this depends on the pre-specified range of clinically relevant risk thresholds. For example, in the 0 to 0.2 range the net benefit is above 0 for all the curves and thus might still be deemed quite stable. However, in the 0.3 to 0.5 range the curves show net benefit spans above and below, and this instability suggests considerable uncertainty in declaring whether the model has net benefit in this range.

## 5.4 Classification and risk grouping

Clinical utility measured by net benefit is calculated across all individuals and may hide instability in classifications at the individual level. That is, a model may have instability about whether individuals fall above or below a particular risk threshold.[29] A similar concern is instability in risk grouping (e.g. classifying individuals into low, medium and high risks), which is often seen in the medical literature, though not something we generally recommend. Nevertheless, for completeness, we note that the bootstrap process can also examine instability in individual classifications. For each individual, we can calculate the proportion of bootstrap models (from step 4) that give a different classification (i.e. above rather than below the threshold, or below rather than above the threshold) than the original model. We refer to this proportion as the *classification instability index*, and it estimates an individual's probability that a different example model would have produced a different classification than the original model. The findings can be shown on a scatter plot, with each individual's classification index (y-axis) plotted against their original predicted value (x-axis). We call this the *classification instability plot*. Ideally the plot should have a very narrow distribution, with index values close to zero for most individuals, except those with predictions close to the risk threshold (as by definition predicted values just below, or just above, the threshold are most vulnerable to changes in classification). However, in small samples the classification instability index may be large (Figure 8), even for those individuals with predictions far away from the threshold (see random forest example in Supplementary Figure S5), which is concerning.

**Figure 8 Instability in decision curves and classification for two models developed using unpenalised logistic regression (see Section 4.1)**

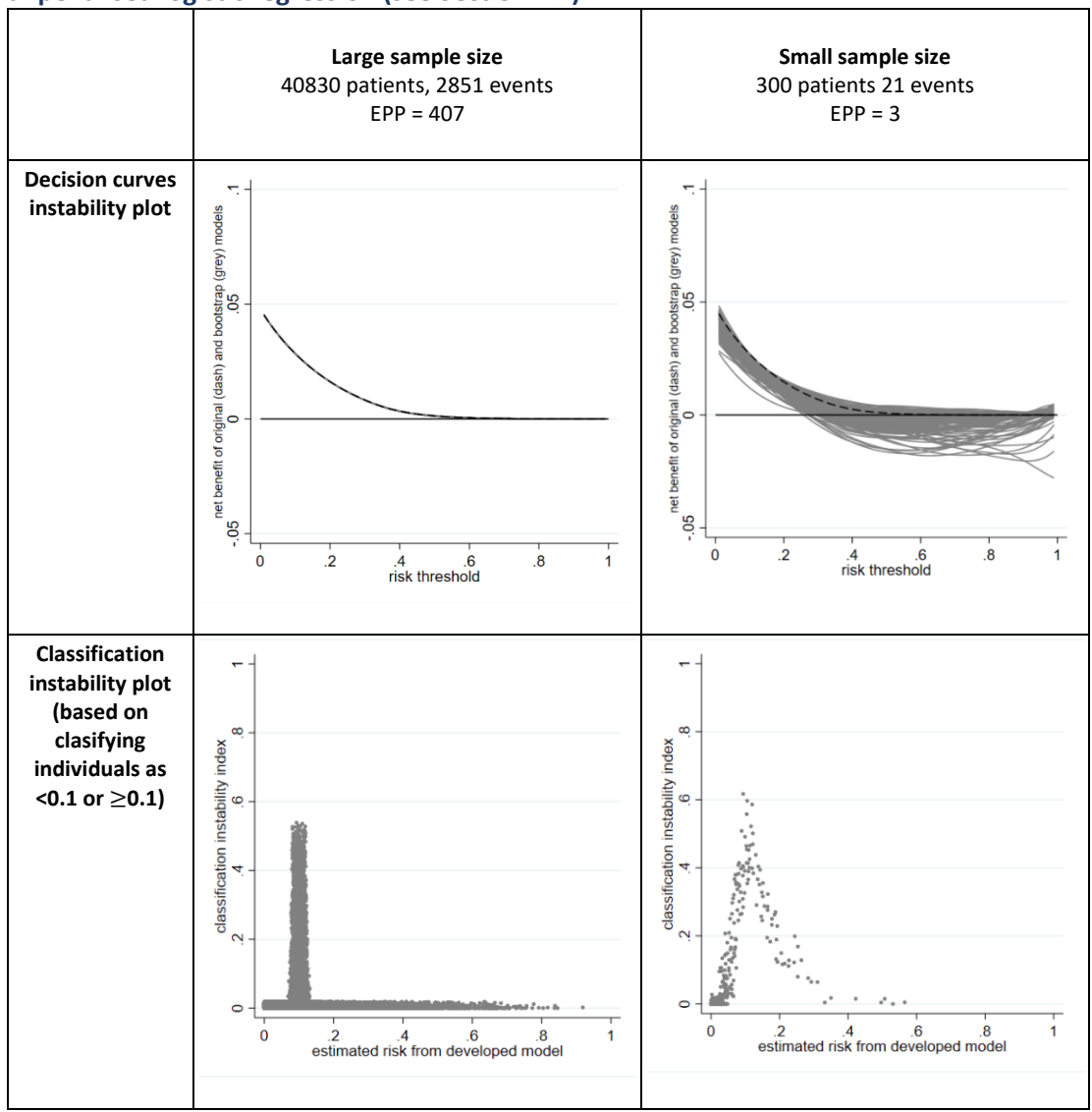

## 5.5 Explainable machine learning and AI models

The black-box nature of machine learning approaches, and more generally artificial intelligence, has raised concerns about implementation of such models in healthcare. To address this, there is a drive toward 'explainable AI'. However, if a model exhibits instability in predictions then any meaningful attempt to 'explain' the model, especially at the individual level, is futile.[44] For example, post-hoc explanation methods such as locally interpretable model-agnostic explanations (LIME) and Shapley values (SHAP) will themselves be unstable and therefore likely misleading.

# 6  Discussion

Clinical prediction models are used to inform individualised decisions about lifestyle behaviours, treatments, monitoring strategies, and other key aspects of healthcare. Hence, it is important they are robust and reliable for all users. Here, we have demonstrated how to examine the (in)stability of a prediction model at the model development stage, using bootstrapping, and proposed various plots and measures to help quantify instability, which we hope will become routinely adopted and presented by model developers. Stata code is provided in the supplementary material for the case study of Section 4, which users can easily adapt for their own models. R code is available at https://github.com/gscollins1973.

Arguably, stability in estimated risks is one of the most important aspects to consider when developing a model. Indeed, in the absence of further validation, the quantification of instability is essential as it exposes the uncertainty and fragility of the newly proposed model. Bootstrapping is an important method for this. Note, though, the bootstrap process only examines instability in the population that the development dataset was sampled from. Indeed, the quality of the bootstrap process is itself dependent on the quality and representativeness of the development sample. Evaluations in other populations require external validation in new data, with sufficient sample size,[45-47] sampled from those other populations.

We defined four levels of stability to be checked, moving from looking at overall or population-level aspects (levels 1 and 2) to the subgroup and individual level. A similar concept is different levels of miscalibration, as considered in the calibration hierarchy of Van Calster et al.[48] Our view is that, at the bare minimum, a model should demonstrate stability at levels 1 and 2. For this reason, we have previously suggested the minimum sample size (and number of predictor parameters) for model development should target precise estimation of the overall risk in the population, low overfitting and small average MAPE.[18-21] However, given that the models are used to guide individuals, there is a strong argument that a model should also demonstrate stability at levels 3 and 4. This requires very large sample sizes, in particular to precisely estimate the effect of key predictors.[18-20] We recognise this may not always be achievable, for example in clinical situations with rare outcomes or when prospective studies are expensive and time-consuming. Data sharing and individual participant data meta-analysis may help address this,[49] but regardless of sample size, stability checks should always be undertaken and reported.

Depending on the intended role of the model, ensuring stability of predictions across the entire range (0 to 1) may be too stringent. In particular, we may desire greatest stability in regions of risk relevant to clinical decision making and be willing to accept lower stability in other regions where miscalibration is less important. For example, for recurrence of venous thromboembolism,[50] predicted risks between about 0.03 and 0.20 have been suggested to warrant clinical action, such as remaining on anticoagulation therapy. Hence, slight to moderate instability in ranges of highest risk (0.5 to 1) is potentially acceptable in this context, as it is unlikely to change decisions that are made based on thresholds defined by low risks (0 to 0.2, say). For this reason, examining the volatility in decision curves after model development can also be helpful, as shown in Section 5. Nevertheless, there is still a strong argument that risk thresholds vary across individuals and settings, and so focusing only on a narrow range of thresholds (e.g., 0 to 0.2) is subjective and an incomplete picture, compared to ensuring stability across the entire spectrum of risks from 0 to 1.

Much of this work has strong synergy to the issue of quantifying uncertainty in prediction models. In particular, the prediction instability plots provide 95% intervals, displaying the 95% range of estimated predictions for individuals across the bootstrap example models. However, we refrain from referring to these as "95% confidence intervals" for estimated risks, as we do not think such terminology is appropriate. Rather the 95% intervals are best viewed as an instability range for predictions from the *particular* model developed. Confidence intervals imply some truth is likely to be within them, but an individual's 'true' risk is difficult to postulate in reality, as no fitted model is true. But we should be aiming for any developed model to at least be internally valid (for the chosen model development approach), and thus demonstrate some degree of stability.

In summary, we strongly encourage researchers to pay close attention to the stability of their models after development and recommend they routinely presenting instability plots and measures, to guide stakeholders (including patients and healthcare professionals) about whether a model is likely (or unlikely) to be reliable in new individuals from the development population, and to help systematic reviewers and peer reviewers critically appraise a model (e.g., in regards to risk of bias classifications [51,52]).

Reference List

1. Riley RD, van der Windt D, Croft P, et al., editors. *Prognosis Research in Healthcare: Concepts, Methods and Impact*. Oxford, UK: Oxford University Press, 2019.

2. Steyerberg EW, Moons KG, van der Windt DA, et al. Prognosis Research Strategy (PROGRESS) 3: prognostic model research. *PLoS Med* 2013;10(2):e1001381.

3. Steyerberg EW. Clinical prediction models: a practical approach to development, validation, and updating (Second Edition). New York: Springer 2019.

4. Nashef SA, Roques F, Michel P, et al. European system for cardiac operative risk evaluation (EuroSCORE). *Eur J Cardiothorac Surg* 1999;16(1):9-13.

5. Nashef SAM, Roques F, Sharples LD, et al. EuroSCORE II. *European Journal of Cardio-Thoracic Surgery* 2012;41(4):734-45.

6. Haybittle JL, Blamey RW, Elston CW, et al. A prognostic index in primary breast cancer. *Br J Cancer* 1982;45(3):361-6.

7. Galea MH, Blamey RW, Elston CE, et al. The Nottingham Prognostic Index in primary breast cancer. *Breast Cancer Res Treat* 1992;22(3):207-19.

8. Dhiman P, Ma J, Andaur Navarro CL, et al. Methodological conduct of prognostic prediction models developed using machine learning in oncology: a systematic review. *BMC Med Res Methodol* 2022;22(1):101.

9. Andaur Navarro CL, Damen JAA, Takada T, et al. Risk of bias in studies on prediction models developed using supervised machine learning techniques: systematic review. *BMJ* 2021;375:n2281.

10. Wynants L, Van Calster B, Collins GS, et al. Prediction models for diagnosis and prognosis of covid-19: systematic review and critical appraisal. *BMJ* 2020;369:m1328.

11. Altman DG, Andersen PK. Bootstrap investigation of the stability of a Cox regression model. *Stat Med* 1989;8(7):771-83.

12. Sauerbrei W, Buchholz A, Boulesteix AL, et al. On stability issues in deriving multivariable regression models. *Biom J* 2015;57(4):531-55.

13. Sauerbrei W, Boulesteix AL, Binder H. Stability investigations of multivariable regression models derived from low- and high-dimensional data. *J Biopharm Stat* 2011;21(6):1206-31.

14. Royston P, Sauerbrei W. Stability of multivariable fractional polynomial models with selection of variables and transformations: a bootstrap investigation. *Stat Med* 2003;22(4):639-59.

15. Royston P, Sauerbrei W. Bootstrap Assessment of the Stability of Multivariable Models. *The Stata Journal* 2009;9(4):547-70.

16. Harrell FE, Jr. Regression Modeling Strategies: With Applications to Linear Models, Logistic and Ordinal Regression, and Survival Analysis (Second Edition). New York: Springer 2015.

17. Riley RD, Van Calster B, Collins GS. A note on estimating the Cox-Snell R(2) from a reported C statistic (AUROC) to inform sample size calculations for developing a prediction model with a binary outcome. *Stat Med* 2021;40(4):859-64.

18. Riley RD, Ensor J, Snell KIE, et al. Calculating the sample size required for developing a clinical prediction model. *BMJ* 2020;368:m441.

19. Riley RD, Snell KI, Ensor J, et al. Minimum sample size for developing a multivariable prediction model: Part II - binary and time-to-event outcomes. *Stat Med* 2019;38(7):1276-96.

20. Riley RD, Snell KIE, Ensor J, et al. Minimum sample size for developing a multivariable prediction model: Part I - Continuous outcomes. *Stat Med* 2019;38(7):1262-75.

21. van Smeden M, Moons KG, de Groot JA, et al. Sample size for binary logistic prediction models: Beyond events per variable criteria. *Stat Methods Med Res* 2019;28(8):2455-74.

22. Sinkovec H, Heinze G, Blagus R, et al. To tune or not to tune, a case study of ridge logistic regression in small or sparse datasets. *BMC Med Res Methodol* 2021;21(1):199.

23. Van Calster B, van Smeden M, De Cock B, et al. Regression shrinkage methods for clinical prediction models do not guarantee improved performance: Simulation study. *Stat Methods Med Res* 2020;29(11):3166-78.

24. Riley RD, Snell KIE, Martin GP, et al. Penalization and shrinkage methods produced unreliable clinical prediction models especially when sample size was small. *J Clin Epidemiol* 2021;132:88-96.

25. Van Houwelingen JC. Shrinkage and penalized likelihood as methods to improve predictive accuracy. *Statistica Neerlandica* 2001;55:17-34.

26. Martin GP, Riley RD, Collins GS, et al. Developing clinical prediction models when adhering to minimum sample size recommendations: The importance of quantifying bootstrap variability in tuning parameters and predictive performance. *Statistical Methods in Medical Research* 2021;30(12):2545-61.

27. Roberts S, Nowak G. Stabilizing the lasso against cross-validation variability. *Computational Statistics & Data Analysis* 2014;70:198-211.

28. Breiman L. Heuristics of Instability and Stabilization in Model Selection. *The Annals of Statistics* 1996;24(6):2350-83.

29. Pate A, Emsley R, Sperrin M, et al. Impact of sample size on the stability of risk scores from clinical prediction models: a case study in cardiovascular disease. *Diagn Progn Res* 2020;4:14.

30. Cleveland WS. Robust Locally Weighted Regression and Smoothing Scatterplots. *Journal of the American Statistical Association* 1979;74(368):829-36.

31. The GUSTO Investigators. An International Randomized Trial Comparing Four Thrombolytic Strategies for Acute Myocardial Infarction. *New England Journal of Medicine* 1993;329(10):673-82.

32. Van Houwelingen JC, Le Cessie S. Predictive value of statistical models. *Stat Med* 1990;9(11):1303-25.

33. Steyerberg EW, Eijkemans WJC, Harrell FE, et al. Prognostic modelling with logistic regression analysis: a comparison of selection and estimation methods in small data sets. *Stat Med* 2000;19::1059-79.

34. Dankowski T, Ziegler A. Calibrating random forests for probability estimation. *Stat Med* 2016;35(22):3949-60.

35. Platt J. Probabilistic Outputs for Support Vector Machines and Comparisons to Regularized Likelihood Methods. *Adv Large Margin Classif* 2000;10

36. van der Ploeg T, Austin PC, Steyerberg EW. Modern modelling techniques are data hungry: a simulation study for predicting dichotomous endpoints. *BMC Medical Research Methodology* 2014;14(1):137.

37. Grote T, Keeling G. On Algorithmic Fairness in Medical Practice. *Camb Q Healthc Ethics* 2022;31(1):83-94.

38. Localio A, Goodman S. Beyond the usual prediction accuracy metrics: Reporting results for clinical decision making. *Annals of Internal Medicine* 2012;157(4):294-95.

39. Moons KG, Altman DG, Vergouwe Y, et al. Prognosis and prognostic research: application and impact of prognostic models in clinical practice. *Bmj* 2009;338:b606.

40. Reilly BM, Evans AT. Translating clinical research into clinical practice: impact of using prediction rules to make decisions. *Ann Intern Med* 2006;144(3):201-9.

41. Vickers AJ, Van Calster B, Steyerberg EW. Net benefit approaches to the evaluation of prediction models, molecular markers, and diagnostic tests. *BMJ* 2016;352:i6.

42. Vickers AJ, Elkin EB. Decision curve analysis: a novel method for evaluating prediction models. *Med Decis Making* 2006;26(6):565-74.

43. Vickers AJ, Cronin AM, Elkin EB, et al. Extensions to decision curve analysis, a novel method for evaluating diagnostic tests, prediction models and molecular markers. *BMC Med Inform Decis Mak* 2008;8:53.

44. Ghassemi M, Oakden-Rayner L, Beam AL. The false hope of current approaches to explainable artificial intelligence in health care. *Lancet Digit Health* 2021;3(11):e745-e50.

45. Archer L, Snell KIE, Ensor J, et al. Minimum sample size for external validation of a clinical prediction model with a continuous outcome. *Stat Med* 2021;40(1):133-46.

46. Riley RD, Debray TPA, Collins GS, et al. Minimum sample size for external validation of a clinical prediction model with a binary outcome. *Stat Med* 2021;40(19):4230-51.

47. Riley RD, Collins GS, Ensor J, et al. Minimum sample size calculations for external validation of a clinical prediction model with a time-to-event outcome. *Stat Med* 2022;41(7):1280-95.

48. Van Calster B, Nieboer D, Vergouwe Y, et al. A calibration hierarchy for risk models was defined: from utopia to empirical data. *J Clin Epidemiol* 2016; 74:167-76.

49. Riley RD, Tierney JF, Stewart LA, editors. *Individual Participant Data Meta-Analysis: A Handbook for Healthcare Research*. Chichester, West Sussex: Wiley, 2021.

50. Ensor J, Riley RD, Moore D, et al. Systematic review of prognostic models for recurrent venous thromboembolism (VTE) post-treatment of first unprovoked VTE. *BMJ Open* 2016;6(5):e011190.

51. Wolff RF, Moons KGM, Riley RD, et al. PROBAST: A Tool to Assess the Risk of Bias and Applicability of Prediction Model Studies. *Ann Intern Med* 2019;170(1):51-58.

52. Collins GS, Dhiman P, Andaur Navarro CL, et al. Protocol for development of a reporting guideline (TRIPOD-AI) and risk of bias tool (PROBAST-AI) for diagnostic and prognostic prediction model studies based on artificial intelligence. *BMJ Open* 2021;11(7):e048008.